\documentclass[a4paper,11pt]{article}
\usepackage[margin=3cm]{geometry}
\usepackage{natbib}
\usepackage{amsfonts,amsmath,amssymb,stmaryrd}
\usepackage{graphicx}
\usepackage{verbatim}
\usepackage{float} 
\usepackage[pdfauthor={David Monniaux},
  pdftitle={Automatic Modular Abstractions for Linear Constraints},
  pdfkeywords={quantifier elimination, abstract interpretation, program transformation, linear inequalities},
  pdfsubject={(POPL '09) algorithms, languages, theory, verification}]{hyperref}
\usepackage{algorithm,algorithmic}

\title{Automatic Modular Abstractions for Linear Constraints}
\date{June 27, 2008}
\author{David Monniaux\\
VERIMAG\thanks{VERIMAG is a joint laboratory of CNRS, Universit\'e Joseph Fourier and Grenoble-INP.}}

\newcommand{\doi}[1]{DOI: \href{http://dx.doi.org/#1}{#1}}

\begin{document}

\maketitle

\begin{abstract}
We propose a method for automatically generating abstract transformers for static analysis by abstract interpretation. The method focuses on linear constraints on programs operating on rational, real or floating-point variables and containing linear assignments and tests.

In addition to loop-free code, the same method also applies for obtaining least fixed points as functions of the precondition, which permits the analysis of loops and recursive functions. Our algorithms are based on new quantifier elimination and symbolic manipulation techniques.

Given the specification of an abstract domain, and a program block, our method automatically outputs an implementation of the corresponding abstract transformer. It is thus a form of program transformation.

The motivation of our work is data-flow synchronous programming languages, used for building control-command embedded systems, but it also applies to imperative and functional programming.
\end{abstract}

\newcommand{\sem}[1]{\llbracket #1 \rrbracket}

\newcommand{\bbQ}{\mathbb{Q}}
\newcommand{\bbN}{\mathbb{N}}
\newcommand{\bbZ}{\mathbb{Z}}
\newcommand{\bbR}{\mathbb{R}}
\newcommand{\parts}[1]{\mathcal{P}(#1)}
\newcommand{\definedAs}{\stackrel{\vartriangle}{=}}

\newcommand{\eabs}{\varepsilon_{\text{abs}}}
\newcommand{\elast}{\varepsilon_{\text{last}}}
\newcommand{\erel}{\varepsilon_{\text{rel}}}

\newcommand{\lfp}{\text{lfp~}}

\section{Introduction}
In program analysis, it is often necessary to prove or infer numerical properties of programs, for instance, in order to prove certain relationships between array indices, or to prove the absence of overflows.
Static program analysis by abstract interpretation obtains properties of variables, or of relationships between variables, representable in an \emph{abstract domain}. Examples of ``classical'' numerical abstract domains for numerical properties include intervals \cite{CousotCousot76-1} --- to each variable $x$ one attaches an interval $[x_{\min}, x_{\max}]$ --- and convex polyhedra \cite{CousotHalbwachs78} --- conjunctions of inequalities $a_1 x_1 + \dots + a_n x_n \leq c$ are inferred.

For each implemented numerical domain and each program instruction, the static analyzer must provide an abstract transfer function, which maps the property before the instruction to a safe property after the instruction (for forward analysis; the reverse is true of backward analysis). For instance, over the intervals, \texttt{z=x+y} is optimally abstracted as $z_{\max}=x_{\max}+y_{\max}$ and $z_{\min}=x_{\min}+y_{\min}$; the transfer functions for polyhedra are more complex. While the designers of abstract interpreters generally strive so that the output property is ``optimal'' (the interval $[z_{\min}, z_{\max}]$ defined above is the least possible one for the inclusion ordering), optimality is not preserved by composition. Consider, for instance,
\texttt{y=x; z=x-y;}
with the precondition that $\texttt{x} \in [0,1]$. The interval for \texttt{z}, obtained from those for \texttt{x} and \texttt{y} by applying the rules of interval arithmetics, is $[-1,1]$; yet, the optimal interval is $\{0\}$. The reason for this loss of precision is that while the computation of the interval for \texttt{z} from those for \texttt{x} and \texttt{y} is locally optimal, it does not take into account the relationship between \texttt{x} and~\texttt{y}.

 Our initial target application was programs written in synchronous data-flow languages such as \textsc{Lustre} \cite{LUSTRE}, \textsc{Simulink} or \textsc{Scade} \cite{Caspi_et_al_2003}. In these languages, operators are built out of elementary operators, introducing many intermediate variables. Successions of small elementary operations may also occur when analyzing low-level code, e.g. assembly \cite{DBLP:conf/cav/GopanR07,DBLP:conf/cc/BalakrishnanR04} or Java bytecode, and they hamper certain static analysis methods due to the reduced size of the code window used for transfer functions \cite{DBLP:conf/cc/LogozzoF08}. Analyzing floating-point code at the assembly level may actually be easier than analyzing higher-level programs, since the semantics of elementary floating-point operations are usually fairly well-defined while the definition and compiling processes of higher-level languages may leave significant leeway \cite{Monniaux_TOPLAS08}. It is therefore important, for such applications, to be able to analyze program blocks as a whole and not as a succession of independent operations.

In the above simple example, we could obtain better precision by using a relational abstract domain linking the inputs and the outputs of the procedure. In general, though, the code fragment may contain tests and loops (or, more generally, semantic fixed points), which complicates the matter (see Sec.~\ref{part:rate_limiter} for a short example whose semantics involves a fixed point).

Ideally, for better precision, the analyzer should provide a (hopefully optimal) abstract transfer function for each possible program block (fragment of code without loops). However, the designers of the analyzer cannot include a hand-coded function for each possible program block to be analyzed, if only because the number of possible program blocks is infinite. Also, the user might want to use abstract domains not pre-programmed in the analyzer. We would like that abstract transfer functions be obtained automatically from the definition of the abstract domain and the source code (or semantics) of the program block.

In this article, we show how to automatically transform program blocks without loops into an effective implementation of their \emph{optimal abstract transfer function}. This optimal transformer maps constraints on the block inputs to the tightest possible constraints on the block output.
This transformation is parametric in the abstract domain used: it takes as inputs both the program block and a specification of the abstract domain, and outputs the corresponding transfer function. The same method applies for both forward and backward analysis by abstract interpretation, though, for the sake of simplicity, the article focuses on forward analysis.

For short, our analysis considers the exact transition relation of loop-free program fragments as an existentially quantified formula. From that formula, it is able to compute the optimal abstract transformer for the fragment with respect to a user-specified abstract domain, or even for the least invariant of the fragment in that abstract domain. The user may specify any abstract domain in the wide class of \emph{template linear abstract domains}~\cite{Colon_CAV03}.

Our method is based upon \emph{quantifier elimination} in the theory of rational linear arithmetic. It has long been known that this theory admitted quantifier elimination, but algorithms remained mostly impractical. Recent improvements in SAT/SMT solving techniques have made it possible to perform quantifier elimination on larger formulas~\cite{Monniaux_LPAR08}.

We also show how to obtain transfer functions for loops, which are also optimal in a certain sense (they compute the least inductive invariant representable in the abstract domain).\medskip

In the beginning of the article, we focus on simple forward analysis of loop-free blocks, then single loops (or single fixed points), for programs dealing with real or rational variables. The same methods apply to integer variables, at the expense of some added abstraction. We show in later sections how to deal with various constructions, including nested loops and arbitrary control-flow graphs, recursive procedures and \emph{floating-point computations}. Our focus was indeed, originally, synchronous data-flow programs operating over real (for modeling) or floating-point (for execution) variables, but we realized that the same technique could apply to a wider spectrum of languages.

Our analysis goes further than most constraint-based static analysis \cite{Sankaranarayana+others/05/Scalable,Sankaranarayanan_SAS04} in that it computes the general form of the optimal postcondition or least inductive invariant as a function of the precondition parameters, not just for specific values of those parameters. For a simple example, if the procedure is invoked on the interval domain and the $z:=x+y$ operation, our transformation outputs $z_{\min}: = x_{\min} + y_{\min}$ and $z_{\max} := x_{\max} + y_{\max}$. This is especially important since the function mapping the input parameters to the output parameters may be non convex (a simple example is the abstraction of the absolute value with respect to intervals from Sec.~\ref{part:template_linear_constraints_transformers}).

In the above case, the abstract transfer function is linear, but in general it is only piecewise linear. It can be expressed as a simple executable program, consisting only of tests and assignments (see an example at the end of Sec.~\ref{part:template_linear_constraints_transformers}). The analysis thus amounts to a \emph{program transformation} from the concrete to the abstract program. An advantage of obtaining the abstract transfer functions in such a form is that it can be compiled as an ordinary program and loaded back into the analyzer for maximal efficiency. The abstract transfer function obtained by the analysis of a block may be retained for future use, since it is valid in any context. An application of our transformation is therefore \emph{modular interprocedural analysis}.

We have so far considered analyzes where the constraints apply to program variables at a given control point. It is also possible to consider relationships between variables at two different control points, especially the entry and exit of procedures. This way, we can also analyze programs with recursive procedures, including the famous McCarthy~91 function~\cite{Manna_McCarthy_69,Manna_Pnueli_JACM70}.

Contrary to most analyzes of numerical properties based on abstract interpretation, our analysis for loops does not use \emph{widening operators} for finding over-approximations of least fixed points. For instance, the set of reachable states at the start of a loop (a \emph{loop invariant}) is expressed as the least fixed point of the transition relation that contains the input precondition. In widening-based analyzes, over-approximations of the set of reachable states after 1, 2, 3, etc. loop iterations are computed, and the analyzer tries to extrapolate these results in order to obtain some ``candidate'' for being a loop invariant. For instance, an abstract analyzer based on intervals may obtain $[1, 2]$, $[1, 3]$, $[1, 5]$, and, because the lower bound of the interval stays stable and the upper bound is unstable, may try $[1,+\infty[$. If $[1,+\infty[$ is stable under the transition relation, then it is a safe invariant, otherwise further widening is needed. Widenings are a major source of imprecision in many static analyzers and their design is somewhat of a ``black art''. While the soundness of the transition relation and the stability test ensure that the analysis results are correct, and the correct construction of the widening operator ensures termination, the quality of the over-approximation obtained (whether it is close to the actual least invariant or far from it) depends on various factors. In contrast, our method is guaranteed to yield least inductive invariants.

In Sec.~\ref{part:linear_formulas}, we recall facts of formulas built out of linear inequalities. In Sec.~\ref{part:template_linear_constraints} we define the class of abstract domains that we consider. In Sec.~\ref{part:template_linear_constraints_transformers}, we show how we obtain optimal abstract transformers as logical formulas, and in Sec.~\ref{part:template_linear_constraints_gen} how to compile these formulas into executable functions. In Sec.~\ref{part:template_linear_constraints_invariants} we show how the same process applies to least inductive invariants. In Sec.~\ref{part:extensions} we show how to deal with various extensions to the admissible domains and operations: how to allow infinite values for constraint parameters, how to allow some class of non-convex domains, how to partition the state space, and how to model floating-point computations using real numbers. In Sec.~\ref{part:control-flow} we shall see how to deal with recursive procedures and arbitrary control-flow graphs.

\section{Linear formulas}
\label{part:linear_formulas}
We consider logical formulas built out of linear inequalities. A \emph{linear expression} is a sum $a_1 v_1 + \dots + a_n v_n$ where the $a_i \in \bbQ$ and the $v_i$ are \emph{variables}. $\bbQ$ denotes the field of rational numbers, $\bbR$ the field of real numbers. A linear inequality is of the form $l > 0$ or $l \geq 0$, where $l$ is a linear expression. Linear inequalities can always be scaled so that they use only integer coefficients, as opposed to rationals. $a \leq b \leq c$ is shorthand for $a \leq b \land b \leq c$. Unquantified formulas are built out of atomic formulas (linear inequalities) using logical connectives $\land$ and~$\lor$. $l = 0$ means $l \geq 0 \land l \leq 0$. A formula is said to be in \emph{disjunctive normal form} (DNF) if it is written as a disjunction $C_1 \lor \dots \lor C_n$, where each of the $C_i$ is a conjunction $A_{i,1} \land \dots \land A_{i,n_j}$ where the $A_{i,j}$ are atomic formulas or negations thereof.
Quantified formulas are built out of the same, plus the universal and existential quantifiers $\forall$ and $\exists$.

The $\bbQ$-models (respectively, $\bbR$-models) of a formula $F$ are mappings $m$ from the free variables of~$F$ to $\bbQ$ (respectively, $\bbR$) such that $m$ verifies the formula; we then note $m \models F$. $F$ is said to be \emph{true} if every assignment is a model (a model is a mapping from the set of variables to $\bbQ$ or $\bbR$), \emph{satisfiable} if it has a model, and \emph{false} or \emph{unsatisfiable} otherwise. Truth and satisfiability are equivalent if $F$ has no free variables.

We say that two formulas $F$ and $G$ with the same free variables are \emph{equivalent}, noted $F \equiv G$, if they have the same models. Any formula is equivalent to a formula in disjunctive normal form, which can be obtained by repeated application of distributivity: $a \land (b \lor c) \equiv (a \land b) \lor (a \land c)$. 
$F$ is said to imply $G$, noted $F \Rrightarrow G$, if all models of $F$ are models of~$G$.
We say that $F$ and $G$ are equivalent \emph{modulo} assumptions $T$, noted $F \equiv_T G$, if $F \land T \equiv G \land T$; we define similarly $F \Rrightarrow_T G$ as $F \land T \Rrightarrow G \land T$. Equivalences modulo assumptions are often used when simplifying formulas. For instance, if we know that a certain program is always used in a context where $T \definedAs a < b$ holds, and program analysis, at some point, generates the formula $F \definedAs \exists x~ a \leq x \leq b$, then this formula can be simplified to $G \definedAs \textsf{true}$.

The theory of linear inequalities admits \emph{quantifier elimination}: for any formula $F$ with quantifiers, there exists a formula $G$ without quantifiers such that $G \equiv F$. There exist several algorithms that compute such a $G$ from $F$. \citet{FerranteRackoff75} proposed a doubly exponential method \cite[Sec.~ 7.3]{BradleyManna07}, which is too slow in practice; we have since proposed another algorithm that takes advantage of the recent improvements in satisfiability testing technology.~\cite{Monniaux_LPAR08} Our algorithm also allows conversion to disjunctive normal form, and formula simplification modulo assumptions.

\section{Optimal Abstraction over Template Linear Constraint Domains}
\label{part:template_linear_constraints_abstraction}
\subsection{Template Linear Constraint Domains}
\label{part:template_linear_constraints}
Let $F$ be a formula over linear inequalities. We call $F$ a domain definition formula if the free variables of $F$ split into $n$ \emph{parameters} $p_1,\dots,p_n$ and $m$ \emph{state variables} $s_1,\dots,s_m$. We note $\gamma_F: \bbQ^n \rightarrow \parts{\bbQ^m}$ defined by $\gamma_F(\vec{p}) = \{ \vec{s} \in \bbQ^m \mid (\vec{p}, \vec{s}) \models F \}$. As an example, the interval abstract domain for 3 program variables $s_1, s_2, s_3$ uses 6 parameters $m_1, M_1, m_2, M_2, m_3, M_3$; the formula is $m_1 \leq s_1 \leq M_1 \land m_2 \leq s_2 \leq M_2 \land m_3 \leq s_3 \leq M_3$.

In this section, we focus on the case where $F$ is a conjunction $L_1(s_1, \dots, s_m) \leq p_1 \land \dots \land L_n(s_1, \dots, s_m) \leq p_n$ of linear inequalities whose left-hand side is fixed and the right-hand sides are parameters. Such conjunctions define the class of \emph{template linear constraint domains}~\cite{Colon_CAV03}. Particular examples of abstract domains in this class are:
\begin{itemize}
\item the intervals (for any variable $s$, consider the linear forms $s$ and $-s$);
\item the difference bound matrices (for any variables $s_1$ and $s_2$, consider the linear form $s_1-s_2$);
\item the octagon abstract domain (for any variables $s_1$ and $s_2$, distinct or not, consider the linear forms $\pm s_1 \pm s_2$)~\cite{Mine_AST_WCRE01}
\item the octahedra (for any tuple of variables $s_1, \dots, s_n$, consider the linear forms $\pm s_1 \dots \pm s_n$).~\cite{Clariso_Cortadella_SAS2004}
\end{itemize}

Remark that $\gamma_F$ is in general not injective, and thus one should distinguish the \emph{syntax} of the values of the abstract domain (the vector of parameters~$\vec{p}$) and their \emph{semantics} $\gamma_F(\vec{p})$. As an example, if one takes $F$ to be $s_1 \leq p_1 \land s_2 \leq p_2 \land s_1+s_2 \leq p_3$, then both $(p_1,p_2,p_3)=(1, 1, 2)$ and $(1, 1, 3)$ define the same set for state variables $s_1$ and $s_2$. If $\vec{u} \leq \vec{v}$ coordinate-wise, then $\gamma_F(\vec{u}) \subseteq \gamma_F(\vec{v})$, but the converse is not true due to the non-uniqueness of the syntactic form.

Take any nonempty set of states $W \subseteq \bbQ^m$. Take for all $i=1,\dots,m$: $p_i = \sup_{\vec{s} \in W} L_i(\vec{s})$. Clearly, $W \subseteq \gamma_F(p_1, \dots, p_m)$, and in fact $\vec{p}$ is such that $\gamma_F(\vec{p})$ is the least solution to this inclusion.
$p_i$ belongs in general to $\bbR \cup \{ +\infty \}$, not necessarily to $\bbQ \cup \{ +\infty \}$. (for instance, if $W = \{ s_1 \mid s_1^2 \leq 2 \}$ and $L_1=s_1$, then $p_1 = \sqrt{2}$). We have therefore defined an $\alpha_F: \parts{\bbR^m} \rightarrow \{ \bot \} \cup (\bbR \cup \{ +\infty \})^n$, and $(\alpha_F, \gamma_F)$ form a \emph{Galois connection}: $\alpha_F$ maps any set to its best upper-approximation. The fixed points of $\alpha_F \circ \gamma_F$ are the \emph{normal forms}. For instance, $s_1 \leq 1 \land s_2 \leq 1 \land s_1 + s_2 \leq 2$ is in normal form, while $s_1 \leq 1 \land s_2 \leq 1 \land s_1 + s_2 \leq 3$ is not.

\subsection{Optimal Abstract Transformers for Program Semantics}
\label{part:template_linear_constraints_transformers}
We shall consider the input-output relationships of programs with rational or real variables. We first narrow the problem to programs without loops and consider programs consisting in linear arithmetic assignments, linear tests, and sequences. 
Noting $a, b, \dots$ the values of program variables $\texttt{a}, \texttt{b}\dots$ at the beginning of execution and $a', b', \dots$ the output values, the semantics of a program $P$ is defined as a formula $\sem{P}$ such that $(a, b, \dots, a', b', \dots) \models P$ if and only if the memory state $(a', b', \dots)$ can be reached at the end of an execution starting in memory state $(a, b, \dots)$:
\begin{description}
\item[Arithmetic]
  $\sem{a:=L(a, b, \dots)+K}_F \definedAs a'=L(a, b, \dots)+K \land b'=b \land c'=c \land \dots$ where $K$ is a real constant and $L$ is a linear form, and $b,c,d\dots$ are all the variables except~$a$;

\item[Tests]
  $\sem{\texttt{if~} c \texttt{~then~} p_1 \texttt{~else~} p_2} \definedAs
   (c \land \sem{p_1}_F) \lor (\neg c \land \sem{p_2}_F)$;

\item[Non deterministic choice]
  $\sem{a:=\texttt{random}} \definedAs b'=b \land c'=c \land \dots$, for all variables except~$a$;

\item[Failure]
  $\sem{\texttt{fail}} \definedAs \textsf{false}$;

\item[Skip]
  $\sem{\texttt{skip}} \definedAs a'=a \land b'=b \land c'=c \land \dots$

\item[Sequence]
$\sem{P_1; P_2}_F \definedAs \exists a'', b'', \dots ~ f_1 \land f_2$ where
$f_1$ is $\sem{P_1}_F$ where $a'$ has been replaced by $a''$, $b'$ by $b''$ etc., 
$f_2$ is $\sem{P_2}_F$ where $a$ has been replaced by $a''$, $b$ by $b''$ etc.
\end{description} 

In addition to  linear inequalities and conjunctions, such formulas contain disjunctions (due to tests and multiple branches) and existential quantifiers (due to sequential composition).

Note that so far, we have represented the concrete denotational semantics \emph{exactly}. This representation of the transition relation using existentially quantified formulas is evidently as expressive as a representation by a disjunction of convex polyhedra (the latter can be obtained from the former by quantifier elimination and conversion to disjunctive normal form), but is more compact in general. This is why we defer quantifier elimination to the point where we compute the abstract transfer relation.

Consider now a domain definition formula $F \definedAs L_1(s_1, s_2, \dots) \leq p_1 \land \dots \land L_n(s_1, s_2, \dots) \leq p_n$ on the program inputs, with parameters $\vec{p}$ and free variables $\vec{s}$, and another $F' \definedAs L'_1(s'_1, s'_2, \dots) \leq p'_1 \land \dots \land L'_n(s'_1, s'_2, \dots) \leq p'_n$ on the program outputs, with parameters $\vec{p'}$ and free variables $\vec{s'}$. Sound forward program analysis consists in deriving a \emph{safe post-condition} from a precondition: starting from any state verifying the precondition, one should end up in the post-condition. Using our notations, the \emph{soundness condition} is written 
\begin{equation}\label{eqn:soundness}
\forall \vec{s},\vec{s'}~ F \land \sem{P} \implies F'
\end{equation}
The free variables of this relation are $\vec{p}$ and $\vec{p'}$: the formula links the value of the parameters of the input constraints to admissible values of the parameters for the output constraints.
Note that this soundness condition can be written as a universally quantified formula, with no quantifier alternation. Alternatively, it can be written as a conjunction of correctness conditions for each output constraint parameter:
$C'_i \definedAs \forall \vec{s},\vec{s'}~ F \land \sem{P} \implies L'_i(\vec{s'}) \leq p'_i.$

Let us take a simple example: if $P$ is the program instruction $z:=x+y$, $F \definedAs x \leq p_1 \land y \leq p_2$, $F' \definedAs z \leq p'_1$, then $\sem{P} \definedAs z' = x+y$, and the soundness condition is
$\forall x, y, z~ (x \leq p_1 \land y \leq p_2 \land z=x+y \implies z \leq p'_1)$. Remark that this soundness condition is equivalent to a formula without quantifiers $p'_1 \geq p_1 + p_2$, which may be obtained through quantifier elimination. Remark also that while any value for $p'_1$ fulfilling this condition is \emph{sound} (for instance, $p'_1=1000$ for $p_1=p_2=1$), only one value is \emph{optimal} ($p'_1=2$ for $p_1=p_2=1$). An optimal value for the output parameter $p'_i$ is defined by $O'_i \definedAs C'_i \land \forall q'_i~ (C'_i[q'_i / p'_i] \implies p'_i \leq q'_i)$. Again, quantifier elimination can be applied; on our simple example, it yields $p'_1 = p_1 + p_2$.

If there are $n$ input constraint parameters $p_1, \dots, p_n$, then the optimal value for each output constraint parameter $p'_i$ is defined by a formula $O'_i$ with $n+1$ free variables  $p_1, \dots, p_n, p'_i$. This formula defines a \emph{partial function} from $\bbQ^n$ to $\bbQ$, in the mathematical sense: for each choice of $p_1, \dots, p_n$, there exist at most a single $p'_i$. The values of $p_1, \dots, p_n$ for which there exists a corresponding $p'_i$ make up the \emph{domain of validity} of the abstract transfer function. Indeed, this function is in general not defined everywhere; consider for instance the program:
\begin{verbatim}
if (x >= 10) { y = random; } else { y = 0; }
\end{verbatim}
If $F = x \leq p_1$ and $F' = y \leq p'_1$, then $O'_1 \equiv p_1 < 10 \land p'_1=0$, and the function is defined only for $p_1 < 10$.

At this point, we have a characterization of the optimal abstract transformer corresponding to a program fragment $P$ and the input and output domain definition formulas as $n$ formulas (where $n$ is the number of output parameters) $O'_i$ each defining a function (in the mathematical sense) mapping the input parameters $\vec{p}$ to the output parameter~$p'_i$.

Another example: the absolute value function $y:=|x|$, again with the interval abstract domain. The semantics of the operation is $(x \geq 0 \land y=x) \lor (x < 0 \land y=-x)$; the precondition is $x \in [x_{\min}, x_{\max}]$ and the post-condition is $y \in [y_{\min}, y_{\max}]$.
Acceptable values for $(y_{\min},y_{\max})$ are characterized by
formula
\begin{equation}
G \definedAs
\forall x~x_{\min} \leq x \leq x_{\max} \implies y_{\min} \leq |x| \leq y_{\max}
\end{equation}
The optimal value for $y_{\max}$ is defined by $G \land \forall y'_{\max}~ G[y'_{\max}/y_{\max}] \allowbreak\implies\allowbreak y_{\max} \leq y'_{\max}$. Quantifier elimination over this last formula gives as characterization for the least, optimal, value for $y_{\max}$:
\begin{multline}\label{eqn:abs_dnf}
(x_{\min} + x_{\max} \geq 0 \land y_{\max}=x_{\max}) \lor\\
(x_{\min} + x_{\max} < 0 \land y_{\max} = -x_{\min}).
\end{multline}
We shall see in the next sub-section that such a formula can be automatically compiled into code such as:
\begin{verbatim}
if (xmin + xmax >= 0) {
  ymax = xmax;
} else {
  ymax = -xmin;
}
\end{verbatim}

\subsection{Generation of the Implementation of the Abstract Domain}
\label{part:template_linear_constraints_gen}
Consider formula~\ref{eqn:abs_dnf}, defining an abstract transfer function.
On this disjunctive normal form we see that the function we have defined is \emph{piecewise linear}: several regions of the range of the input parameters are distinguished (here, $x_{\min} + x_{\max} < 0$ and $x_{\min} + x_{\max} \geq 0$), and on each of these regions, the output parameter is a linear function of the input parameters. Given a disjunct (such as $y_{\max} = -x_{\min} \land x_{\min} + x_{\max} < 0$), the domain of validity of the disjunct can be obtained by existential quantifier elimination over the result variable (here $\exists y_{\max}~(y_{\max} = -x_{\min} \land x_{\min} + x_{\max} < 0)$). The union of the domains of validity of the disjuncts is the domain of validity of the full formula. The domains of validity of distinct disjuncts can overlap, but in this case, since $O'_i$ defines a function in the mathematical sense, the functions defined by such disjuncts coincide on their overlapping domains of validity.

This suggests a first algorithm for conversion to an executable form:
\begin{enumerate}
\item Put $O'_i$ into quantifier-free, disjunctive normal form $C_1 \land \dots \land C_n$.
\item For each disjunct $C_i$, obtain the validity domain $V_i$ as a conjunction of linear inequalities and solve for $p'_i$ (obtain $p'_i$ as a linear function $v_i$ of the $p_1, \dots, p_n$).
\item Output the result as a cascade of if-then-else and assignments, as in the example at the end of Sec.~\ref{part:template_linear_constraints_transformers}.
\end{enumerate}

\begin{algorithm}
\caption{$\textsc{ToITEtree}(F,z,T)$: turn a formula defining $z$ as a function of the other free variables of $F$ into a tree of if-then-else constructs, assuming that $T$ holds.}
\label{alg:ToIfThenElseTree}

\begin{algorithmic}
\STATE $D (= C_1 \land \dots \land C_n) \gets \textsc{QElimDNFModulo}(\{\},F,T)$
\FORALL{$C_i \in D$}
  \STATE $P_i \gets \textsc{QElimDNFModulo}(z, F, T)$
\ENDFOR
\STATE $P \gets \textsc{Predicates}(P_1, \dots, P_n)$
\IF{$P = \emptyset$}
  \ENSURE $\exists z~F$ is always true
  \STATE $O \gets \textsc{Solve}(D, z)$
\ELSE
  \STATE $K \gets \textsc{Choose}(P)$
  \STATE $O \gets \textsf{IfThenElse}(K, \textsc{ToITEtree}(F,z,T \land K),\allowbreak \textsc{ToITEtree}(F,z,T \land \neg K))$
\ENDIF
\end{algorithmic}
\end{algorithm}

An if-then-else cascade may be inefficient, since identical conditions may have to be tested several times. We could of course factor out all conditions and assign them to Boolean variables, but then, some of the tests performed may actually not be needed. We therefore propose an algorithm for building an if-then-else \emph{tree}. The idea of the algorithm is as follows:
\begin{itemize}
\item Each path in the if-then-else tree corresponds to a conjunction $C$ of conditions (if one goes through the ``if'' branch of \texttt{if (a)} and the ``else'' branch of \texttt{if (b)}, then the path corresponds to $a \land \neg b$).
\item The formula $O'_i$ is simplified relatively to~$C$, a process that prunes out conditions that are always or never satisfied when $C$~holds.
\item If the path is deep enough, then the simplified formula becomes a conjunction. One then solves this conjunction to obtain the computed variable (here, $y_{\max}$) as a function.
\end{itemize}

Our algorithm $\textsc{ToITEtree}(F,z,T)$ (Alg.~\ref{alg:ToIfThenElseTree}) uses a function $\textsc{QElimDNFModulo}\allowbreak(\vec{v},F,T)$ that, given a possibly empty vector of variables $\vec{v}$, a formula $F$ and a formula $T$, outputs a quantifier-free formula $F'$ in disjunctive normal form such that $F' \equiv_T \exists \vec{v}~F$ and no useless predicates are used. $\textsc{Predicates}(F)$ returns the set of atomic predicates of~$F$. $\textsc{Solve}(D, z)$ solves a minimal disjunction $D$ of inequalities for variable $z$, assuming that there is at most one solution for $z$ for each choice of the other variables; one simple way to do that is to look for any constraint of the form $z \geq L$ or $z \leq L$ and output $z = L$. $\textsc{Choose}(P)$ chooses any predicate in $P$ (one good heuristic seems to be to choose the most frequent in $P_1, \dots, P_n$).

Let us take, as a simple example, formula~\ref{eqn:abs_dnf}. We wish to obtain $y_{\max}$ as a function of $x_{\min}$ and $x_{\max}$, so in the algorithm \textsc{ToITEtree} we set $z \definedAs y_{\max}$. $C_1$ is the first disjunct $x_{\min} + x_{\max} \geq 0 \land y_{\max}=x_{\max}$, $C_2$ is the second disjunct $x_{\min} + x_{\max} < 0 \land y_{\max} = -x_{\min}$. We project $C_1$ and $C_2$ parallel to $y_{\max}$, obtaining respectively $P_1 = (x_{\min}+x_{\max} \geq 0)$ and $P_2 = (x_{\min}+x_{\max} < 0)$. We choose $K$ to be the predicate $x_{\min}+x_{\max} \geq 0$ (in this case, the choice does not matter, since $P_1$ and $P_2$ are the negation of each other).
\begin{itemize}
\item The first recursive call to \textsc{ToITEtree} is made in the context of $T \definedAs (x_{\min}+x_{\max} \geq 0)$. Obviously, $F \land T \equiv (y_{\max}=x_{\max}) \land T$ and thus $(\exists y_{\max} F) \land T \equiv T$.

$\textsc{QElimDNFModulo}(y_{\max}, F, T)$ will then simply output the formula ``true''. It then suffices to solve for $y_{\max}$ in $y_{\max}=x_{\max}$. This yields the formula for computing the correct value of $y_{\max}$ in the cases where $x_{\min}+x_{\max} \geq 0$.

\item The second recursive call is made in the context of  $T \definedAs (x_{\min}+x_{\max} < 0 $. The result is $y_{\max}=-x_{\min}$, the formula for computing the correct value of $y_{\max}$ in the cases where $x_{\min}+x_{\max} < 0$.
\end{itemize}
These two results are then reassembled into an if-then-else statement, yielding the program at the end of \S\ref{part:template_linear_constraints_transformers}.

The algorithm terminates because paths of depth $d$ in the tree of recursive calls correspond to truth assignments to $d$ atomic predicates among those found in the domains of validity of the elements of the disjunctive normal form of $F$. Since there is only a finite number of such predicates, $d$ cannot exceed that number. A single predicate cannot be assigned truth values twice along the same path because the simplification process in $\textsc{QElimDNFModulo}$ erases this predicate from the formula.

\subsection{Least Inductive Invariants}
\label{part:template_linear_constraints_invariants}

\begin{figure}
\begin{center}
\includegraphics[width=0.5\columnwidth]{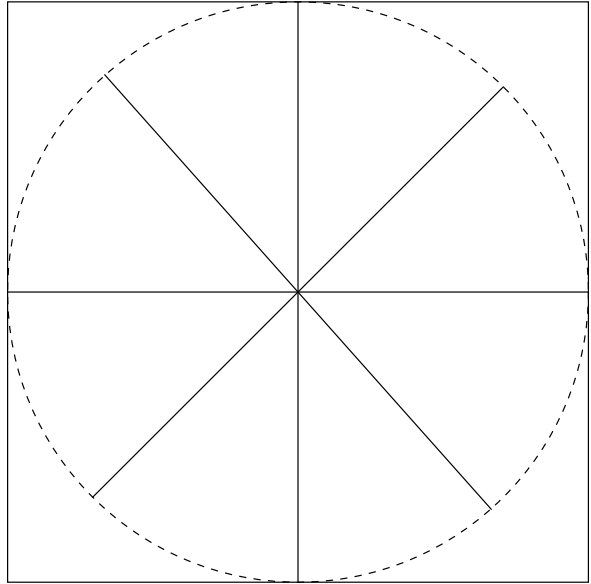}
\end{center}
\caption{The least fixed point representable in the domain ($\lfp(\alpha \circ f \circ \gamma)$) is not necessarily the least approximation of the least fixed point  ($\alpha(\lfp f)$) inside the abstract domain. For instance, if we take a program initialized by $x \in [-1,1]$ and $y=0$, and at each iteration, we rotate the point by 45$^\circ$, the least invariant is an 8-point star, and the best approximation inside the abstract domain of intervals is the square $[-1,1]^2$. However, this square is not an inductive invariant: no rectangle (product of intervals) is stable under the iterations, thus there is no abstract inductive invariant.}
\label{fig:unstable}
\end{figure}

We have so far considered programs without loops. We shall now see that not only can we compute the optimal abstract post-condition of a block as a simple, executable function of the parameters of the precondition, but we can also compute the parameters of the least inductive invariant of a program block that is of the form specified by the abstract domain.%
\footnote{In order to specify the least invariant, we would have to quantify over all sets of states, then filter those which are inductive invariants. This is second-order quantification, which we cannot handle. By restricting ourselves to invariants of a certain shape, we replace it by first order quantification.}
Beware that this least inductive invariant found in the abstract domain is in general different from the least element of the abstract domain that includes the least inductive invariant of the system (Fig.~\ref{fig:unstable}).

\subsubsection{Stability Inequalities}
Consider a program fragment: \verb@while (c) { p; }@.
We have domain definition formulas $F \definedAs L_1(s_1, \dots, s_m) \leq p_1 \land \dots \land L_n(s_1, \dots, s_m) \leq p_n$ for the precondition of the program fragment , and $F' \definedAs L'_1(s_1, \dots, s_m) \leq p'_1 \land \dots \land L'_n(s_1, \dots, s_m) \leq p'_n$ for the invariant.

Define $G = \sem{c} \land \sem{p}$. $G$ is a formula whose free variables are $s_1, \dots, s_m, s'_1, \dots, s'_m$ such that  $(s_1, \dots, s_m, s'_1, \dots, s'_m) \models G$ if and only if the state $(s'_1, \dots, s'_m)$ can be reached from the state $(s_1, \dots, s_m)$ in exactly one iteration of the loop. A set $W \subseteq \bbQ^m$ is said to be an \emph{inductive invariant} for the head of the loop if $\forall \vec{s} \in W, \forall \vec{s'}~(\vec{s}, \vec{s'}) \models G \implies \vec{s'} \in W$. We seek inductive invariants of the shape defined by $F'$, thus solutions for $\vec{p'}$ of the \emph{stability condition}:
\begin{equation}\label{eqn:stability}
\forall \vec{s}, \vec{s'}~ F' \land G \implies F'[\vec{s'}/\vec{s}].
\end{equation}

Not only do we want an inductive invariant, but we also want the initial states of the program to be included in it. The condition then becomes
\begin{equation}
H \definedAs (\forall \vec{s}, F \implies F') \land
(\forall \vec{s}, \vec{s'}~ F' \land G \implies F'[\vec{s'}/\vec{s}])
\end{equation}
This formula links the values of the input constraint parameters $p_1, \dots, p_n$ to acceptable values of the invariant constraint parameters $p'_1, \dots, p'_n$.
In the same way that our soundness or correctness condition on abstract transformers allowed any sound post-condition, whether optimal or not, this formula allows any inductive invariant of the required shape as long as it contains the precondition, not just the least one.

The intersection of sets defined by $\vec{p'}_1$ and $\vec{p'}_2$ is defined by $\min(\vec{p'}_1, \vec{p'}_2)$. More generally, the intersection of a family of sets, unbounded yet closed under intersection, defined by $\vec{p'} \in Z$ is defined by $\min \{ p' \mid p' \in Z\}$.
We take for $Z$ the set of acceptable parameters $\vec{p'}$ such that $\vec{p'}$ defines an inductive invariant and $\forall \vec{s}, F \implies F'$; that is, we consider only inductive invariants that contain the set $I = \{ \vec{s} \mid F \}$ of precondition states.

We deduce that $p'_i$ is uniquely defined by:
$p'_i = \min(\exists p'_1, \allowbreak\dots,\allowbreak p'_{i-1},\allowbreak p'_{i+1},\allowbreak \dots,\allowbreak p'_n~H)$
which can be rewritten as
\begin{equation}
(\exists p'_1, \dots, p'_{i-1}, p'_{i+1}, \dots, p'_n~H) \land
(\forall \vec{q'}~H[\vec{q'}/\vec{p'}] \implies p'_i \leq q'_i)
\end{equation}
The free variables of this formula are $p_1, \dots, p_n, p'_i$. This formula defines a function (in the mathematical sense) defining $p'_i$ from $p_1, \dots, p_n$. As before, this function can be compiled to an executable version using cascades or trees of tests.

\subsubsection{Simple Loop Example}
\label{part:loop_example}
To show how the method operates in practice, let us consider first a very simple example (\texttt{something\_happens} is a nondeterministic choice):
\begin{verbatim}
int i=0;
while (i <= n) {
  if (something_happens) {
    i=i+1;
    if (i == n) {
      i=0;
    }
  }
}
\end{verbatim}

Let us abstract \texttt{i} at the head of the loop using an interval $[i_{\min},i_{\max}]$. For simplicity, we consider the case where the loop is at least entered once, and thus $i=0$ belongs to the invariant.
For better precision, we model each comparison $x \neq y$ over the integers as
$x >= y+1 \lor x <= y-1$; similar transformations apply for other operators.
The formula expressing that such an interval is an inductive invariant is:%
\begin{multline}
i_{\min}\leq 0\land 0\leq
   i_{\max}\land \forall i \forall i'~
   ((i_{\min}\leq i\land i\leq i_{\max}\land\\
   (((i+1\leq n-1\lor i+1\geq
   n+1)\land i'=i+1)\lor\\
   (i+1=n+1\land i'=0)\lor
   i'=i))\implies
   (i_{\min}\leq i'\land
   i'\leq i_{\max}))
\end{multline}

Quantifier elimination produces:
\begin{multline}
(i_{\min}\leq 0\land
   i_{\max}\geq 0\land
   i_{\max}<n\land
   -i_{\min}+n-2<0)\lor\\
   (i_{\min}\leq 0\land
   i_{\max}\geq 0\land
   i_{\max}-n+1\geq 0\land
   i_{\max}<n)
\end{multline}

The formulas defining optimal $i_{\min}$ and $i_{\max}$ are:
\begin{eqnarray}
i_{\min}\geq 0\land
   i_{\min}\leq 0\land n>0\\
(i_{\max}= 0\land \land n>0\land n<2) \lor
(i_{\max} = n-1 \land i_{\max}\geq 1)
\end{eqnarray}

We note that this invariant is only valid for $n > 0$, which is unsurprising given that we specifically looked for invariants containing the precondition $i = 0$. The output abstract transfer function is therefore:
\begin{verbatim}
if (n <= 0) {
  fail();
} else {
  iMin = 0;
  if (n < 2) {
    iMax = 0;
  } else /* n >= 2 */
    iMax = n-1;
  }
}
\end{verbatim}
The case disjunction \texttt{n < 2} looks unnecessary, but is a side effect of the use of rational numbers to model a problem over the integers. The resulting abstract transfer function is optimal, but on such a simple case, one could have obtained the same using polyhedra \cite{CousotHalbwachs78} or octagons \cite{Mine_AST_WCRE01}.

Let us now consider the same program, simply replacing \texttt{n} by the constant \texttt{20}. All implementations of intervals (and thus of octagons and polyhedra, since we only have one variable), will overshoot the $i_{\max}=19$ target when using the traditional widening and narrowing strategies: they will compute $\texttt{i} \in [0,0]$, then $\in [0, 1]$, $\in [0,2]$ and widen to $[0,+\infty[$, and narrowing will not reduce the interval. Even if we replaced \texttt{i == 20} by \texttt{i >= 20}, narrowing would still fail to reduce the interval due to the nondeterministic choice since the concrete transfer function $f$, mapping sets of states at the head of the loop to sets of states at the next iteration, is expansive: for all set of states $W$, $W \subseteq f(W)$. This is a well-known weakness of the widening/narrowing approach, and the workaround is a \emph{syntactic} trick known as \emph{widening up to} or \emph{widening with thresholds}: for all variables, the constants to which it is compared are gathered and used as widening steps \cite[Sec.~7.1.2]{ASTREE_PLDI03}. This syntactic approach fails if tests are more indirect, whereas our semantic approach is not affected.

\subsubsection{Synchronous Data Flow Example: Rate Limiter}
\label{part:rate_limiter}
To go back to the original problem of floating-point data in data-flow languages, let us consider the following library block: a \emph{rate limiter}. When compiled into~C, such a block in inserted in a reactive loop, as shown below, where \verb@assume(c)@ stands for \verb@if (c) {} else {fail();}@:
\begin{verbatim}
while (true) {
  ...
  e1 = random(); assume(e1 >= e1min && e1 <= e1max);
  e2 = random(); assume(e2 >= e2min && e2 <= e2max);
  e3 = random(); assume(e3 >= e3min && e3 <= e3max);
  olds1 = s1;
  if (random) {
    s1 = e3;
  } else {
    if (e1 - olds1 < -e2) {
      s1 = olds1 - e2;
    }
    if (e1 - olds1 > e2) {
      s1 = olds1 + e2;
    }
  }
  ...
}
\end{verbatim}

We are interested in the input-output behavior of that block: obtain bounds on the output \texttt{s1} of the system as functions of bounds on the inputs (\texttt{e1}, \texttt{e2}, \texttt{e3}). Note that in this case, \texttt{s1}, \texttt{e1}, \texttt{e2}, \texttt{e3} are \emph{streams}, not single scalars.
One difficulty is that the \texttt{s1} output is memorized, so as to be used as an input to the next computation step. The semantics of such a block is therefore expressed as a fixed point.

We wish to know the least inductive invariant of the form
${s_1}_{\textrm{min}} \leq s_1 \leq {s_1}_{\textrm{max}}$ under the assumption that ${e_1}_{\textrm{min}} \leq {e_1}_{\textrm{max}} \land
{e_2}_{\textrm{min}} \leq {e_2}_{\textrm{max}} \land
{e_3}_{\textrm{min}} \leq {e_3}_{\textrm{max}}$.
The stability condition yields, after quantifier elimination and projection on ${s_1}_{\textrm{max}}$ the condition ${s_1}_{\textrm{max}} \geq {e_1}_{\textrm{max}} \land {s_1}_{\textrm{max}} \geq {e_3}_{\textrm{max}}$. Minimization then yields an expression that can be compiled to an if-then-else tree:
\begin{verbatim}
if (e1max > e3max) {
  s1max = e1max;
} else {
  s1max = e3max;
} 
\end{verbatim}

This result, automatically obtained, coincides with the intuition that a rate limiter (at least, one implemented with exact arithmetic) should not change the range of the signal that it processes. This program fragment has a rather more complex behavior if all variables and operations are IEEE-754 floating-point, since rounding errors introduce slight differences of regimes between ranges of inputs (Sec.~\ref{part:float}, \ref{part:experiments}). Rounding errors in the program to be analyzed introduce difficulties for analyzes using widenings, since invariant candidates are likely to be ``almost stable'', but not truly stable, because of these errors. Again, there exist workarounds so that widening-based approaches can still operate \cite[Sec.~7.1.4]{ASTREE_PLDI03}.

\section{Extensions to the Admissible Domains and Operations}
\label{part:extensions}
The class of domains and program constructs of the preceding section may seem too limited. We shall see here a few extensions.

\subsection{Infinities}
Consider the interval abstract domain, defined by $x \leq p_2 \land -x \leq p_1$. The techniques explained in Sec.~\ref{part:template_linear_constraints} allow only finite bounds. Yet, it makes sense that $p_1$ and $p_2$ could be equal to $+\infty$ so as to represent infinite intervals. This can be easily achieved by a minor alteration to our definitions. Each parameter $p_i$ is replaced by two parameters $p^b_i$ and $p^\infty_i$. $p^\infty_i$ is constrained to be in $\{0,1\}$ (if the quantifier elimination procedure in use allows Boolean variables, then $p^\infty_i$ can be taken as a Boolean variable); $p^\infty_i=0$ means that $p_i$ is finite and equal to $p_i^b$, $p^\infty_i=1$ means $p_i = +\infty$. $L_i \leq p_i$ becomes $(p^\infty_i > 0) \lor (L_i \leq p^b_i)$, $L_i < p_i$ becomes $(p^\infty_i > 0) \lor (L_i < p^b_i)$. After this rewriting, all formulas are formulas of the theory of linear inequalities without infinities and are amenable to the appropriate algorithms.

\subsection{Non-Convex Domains}
\label{part:non-convex}
Section~\ref{part:template_linear_constraints} constrains formulas to be conjunctions of inequalities of the form $L_i \leq p_i$. What happens if we consider formulas that may contain disjunctions?

The template linear constraint domains of section~\ref{part:template_linear_constraints} have a very important property: they are closed under (infinite) intersection; that is, if we have a family $\vec{p} \in W$, then there exist $p_0$ such that $\bigcap_{\vec{p} \in W} \gamma_F(\vec{p}) = \gamma_F(\vec{p}_0)$ (besides, $p_0 = \inf \{ \vec{p} \mid \vec{p} \in W \}$). This is what enables us to request the \emph{least} element that contains the exact post-condition, or the least inductive invariant in the domain: we take the intersection of all acceptable elements.

Yet, if we allow non-convex domains, there does not necessarily exist a least element $\gamma_F(\vec{p})$ such that $S \subseteq \gamma_F(\vec{p})$. Consider for instance $S = \{0, 1, 2\}$ and $F$ representing unions of two intervals $((-x \leq p_1 \land x \leq p_2) \lor (-x \leq p_3 \land x \leq p_4)) \land p_2 \leq p_3$. There are two, incomparable, minimal elements of the form $\gamma_F(\vec{p})$: $p_1=p_2=0 \land p_3=-1 \land p_4=2$ and $p_1=0 \land p_2=1 \land p_3=-2 \land p_4=2$.

We consider formulas $F$ built out of linear inequalities $L_i(s_1, \allowbreak\dots,\allowbreak s_n) \leq p_i$ as atoms, conjunctions, and disjunctions. 
By induction on the structure of $F$, we can show that $\gamma_F: (\bbR \cup \{ -\infty \})^n \rightarrow \parts{\bbR^n}$ is inf-continuous; that is, for any descending chain $(\vec{p}_i)_{i \in I}$ such that $\lim_i \vec{p}_i = \vec{p}_\infty$, then $\gamma_F(\vec{p}_i)$ is decreasing and $\bigcap_{i \in I} \gamma_F(\vec{p}_i) = \gamma_F(\vec{p}_\infty)$. The property is trivial for atomic formulas, and is conserved by greatest lower bounds ($\land$) as well as binary least upper bounds ($\lor$).

Let us consider a set $S \subseteq \parts{\bbR^n}$, stable under arbitrary intersection (or at least, greatest lower bounds of descending chains). $S$ can be for instance the set of invariants of a relation, or the set of over-approximations of a set~$W$. $\gamma_F^{-1}(S)$ is the set of suitable domain parameters; for instance, it is the set of parameters representing inductive invariants of the shape specified by $F$, or the set of representable over-approximations of~$W$. $\gamma_F^{-1}(S)$ is stable under greatest lower bounds of descending chains: take a descending chain $(\vec{p}_i)_{i \in I}$, then $\gamma_F(\lim_i \vec{p}_i) = \cap_i \gamma_F(\vec{p}_i) \in S$ by inf-continuity and stability of $S$. By Zorn's lemma, $\gamma_F^{-1}(S)$ has at least one minimal element.

Let $P[\vec{p}]$ be a formula representing~$\gamma_F^{-1}(S)$ (Sec.~\ref{part:template_linear_constraints} proposes formulas defining safe post-conditions and inductive invariants). The formula $G[\vec{p}] \definedAs P[\vec{p}] \land \forall \vec{p'}~ P[\vec{p'}] \land \vec{p'} \leq \vec{p} \implies \vec{p} \leq \vec{p'}$ defines the minimal elements of $\gamma^{-1}(S)$.

For instance, consider $\vec{p}=(a,b,c,d)$, $F \definedAs (-x \leq a \land x \leq b) \lor (-x \leq c \land x \leq d)$, representing unions of two intervals $[-a, b] \cup [-c,d]$. We want upper-approximations of the set $\{0, 1, 3\}$; that is $P[\vec{p}] \definedAs \forall x~ (x=0 \lor x=1 \lor x=3 \implies F[\vec{p},x])$. We add the constraint that $-a \leq b \land b \leq -c \land -c \leq d$, so as not to obtain the same solutions twice (by exchange of $(a,b)$ and $(c,d)$) or solutions with empty intervals. By quantifier elimination over $G$, we obtain
$(a = 0 \land b = 1 \land c = -3 \land d=3) \lor (a = 0 \land b = 0 \land c = -1 \land d=3)$, that is, either $[0,0] \cup [1, 3]$ or $[0,1] \cup [3, 3]$.

\subsection{Domain Partitioning}
\label{part:partitioning}
Non-convex domains, in general, are not stable under intersections and thus ``best abstraction'' problems admit multiple solutions as minimal elements of the set of correct abstractions. There are, however, non-convex abstract domains that are stable under intersection and thus admit least elements as well as the template linear constraint domains of Sec.~\ref{part:template_linear_constraints}: those defined by partitioning of the state space. Consider pairwise disjoint subsets $(C_i)_{i \in I}$ of the state space $\bbQ^m$, and abstract domains stable under intersection $(S_i)_{i \in I}$, $S_i \subseteq \parts{C_i}$. Elements of the partitioned abstract domain are unions $\bigcup_{i \in I} s_i$ where $s_i \in S_i$. If $\left(\bigcup_i s_{i,j}]\right)_{j \in J}$ is a family of elements of the domain, then $\bigcap_{j \in J}\left(\bigcup_{i \in I} s_{i,j}]\right) = \bigcup_{i \in I} \bigcap_{j \in J} s_{i,j}$; that is, intersections are taken separately in each~$C_i$.

Take a family $(F_i[\vec{p}])_{i \in I}$ of formulas defining template linear constraint domains (conjunctions of linear inequalities $L_i(s_1,\allowbreak \dots,\allowbreak s_n) \leq p_i$) and a family $(C_i)_{i \in I}$ of formulas such that for all $i$ and $i'$, $C_i \land C_{i'}$ is equivalent to \textsf{false} and $C_1 \lor \dots \lor C_l$ is equivalent to \textsf{true}. $F = (C_1 \land F_1) \lor \dots \lor (C_l \land F_l)$ then defines an an abstract domain such that $\gamma_F$ is a inf-morphism. All the techniques of  Sec.~\ref{part:template_linear_constraints} then apply.

\subsection{Floating-Point Computations}
\label{part:float}
Real-life programs do not operate on real numbers; they operate on fixed-point or floating-point numbers. Floating point operations have few of the good algebraic properties of real operations; yet, they constitute approximations of these real operations, and the \emph{rounding error} introduced can be bounded.

In IEEE floating-point \cite{IEEE-754}, each atomic operation (noting $\oplus$, $\ominus$, $\otimes$, $\oslash$, $\sqrt{}_f$ for operations so as to distinguish them from the operations $+$, $-$, $\times$, $/$, $\sqrt{}$ over the reals) is mathematically defined as the image of the exact operation over the reals by a rounding function.%
\footnote{We leave aside the peculiarities of some implementations, such as those of most C compilers over the 32-bit Intel platform where there are ``extended precisions'' types used for some temporary variables and expressions can undergo double rounding.~\cite{Monniaux_TOPLAS08}}
This rounding function, depending on user choice, maps each real $x$ to the nearest floating-point value $r_n(x)$ (\emph{round to nearest mode}, with some resolution mechanism for non representable values exactly in the middle of two floating-point values), $r_{-\infty}(x)$ the greatest floating-point value less or equal to $x$ (\emph{round toward $-\infty$}), $r_{+\infty}(x)$ the least floating-point value greater or equal to $x$ (\emph{round toward $+\infty$}), $r_0(x)$ the floating-point value of the same sign as $x$ but whose magnitude is the greatest floating-point value less or equal to $|x|$ (\emph{round toward $0$}). If $x$ is too large to be representable, $r(x)=\pm\infty$ depending on the size of $x$

The semantics of the rounding operation cannot be exactly represented inside the theory of linear inequalities.%
\footnote{To be pedantic, since IEEE floating-point formats are of a finite size, the rounding operation could be exactly represented by enumeration of all possible cases; this would anyway be impossible in practice due to the enormous size of such an enumeration.}
As a consequence, we are forced to use an axiomatic over-approximation of that semantics: a formula linking a real number $x$ to its rounded version~$r(x)$.

\citet{Mine_ESOP04} uses an inequality $|r(x)-x| \leq \erel\cdot |x| + \eabs$, where $\erel$ is a \emph{relative error} and $\eabs$ is an \emph{absolute error}, leaving aside the problem of overflows. The relative error is due to rounding at the last binary digit of the significand, while the \emph{absolute error} is due to the fact that the range of exponents is finite and thus that there exists a least positive floating-point number and some nonzero values get rounded to zero instead of incurring a relative error.

Because our language for axioms is richer than the interval linear forms used by Min\'e, we can express more precise properties of floating-point rounding. We recall briefly the characteristics of IEEE-754 floating-point numbers. 
Nonzero floating point numbers are represented as follows:
$x = \pm s.m$ where $1 \leq m < 2$ is the \emph{mantissa} or
\emph{significand}, which
has a fixed number $p$ of bits, and $s =
2^e$ the \emph{scaling factor} ($E_{\min} \leq e \leq E_{\max}$
is the \emph{exponent}).
The difference introduced by changing the last binary digit of the
mantissa is $\pm s.\elast$ where $\elast = 2^{-(p-1)}$:
the \emph{unit in the last place} or \emph{ulp}.
Such a decomposition is unique for a given number
if we impose that the leftmost digit of the
mantissa is $1$ --- this is called a \emph{normalized representation}.
Except in the case of numbers of very small magnitude, IEEE-754 always
works with normalized representations. There exists a least positive normalized number $m_{\textrm{normal}}$ and a least positive denormalized number $m_{\textrm{denormal}}$, and the denormals are the multiples of $m_{\textrm{denormal}}$ less than $m_{\textrm{normal}}$. All representable numbers are multiples of~$m_{\textrm{denormal}}$.

Consider for instance floating-point addition or subtraction $x=\pm a \pm b$. Suppose that $0 \leq x \leq m_{\textrm{normal}}$. $a$ and $b$ are multiples of $m_{\textrm{denormal}}$ and thus $a-b$ is exactly represented as a denormalized number; therefore $r(x)=x$. If $x > m_{\textrm{normal}}$, then $|r(x)-x| \leq \erel.x$. The cases for $x \leq 0$ are symmetrical. We can therefore characterize $r(x)-x$ using linear inequalities through case analysis over $x$: $\textit{Round}_+(a \oplus b, a+b)$ (respectively, $\textit{Round}_+(a \ominus b, a-b)$) holds, where
\begin{multline}
\textit{Round}_+(r,x) \definedAs (x \leq m_{\textrm{normal}} \land r = x)\\
  \lor (x > m_{\textrm{normal}} \land -\erel.x \leq r - x \leq \erel.x
\end{multline}
\begin{multline}
\textit{Round}(r,x) \definedAs (x =0 \land r =0) \lor\\
  (x > 0 \land r \geq 0 \land \textit{Round}_+(r,x)) \lor\\
  (x < 0 \land r \leq 0 \land \textit{Round}_+(-r,-x))
\end{multline}

To each floating-point expression $e$, we associated a ``rounded-off'' variable $r_e$, the value of which we constrain using $\textit{Round}(r_e,e)$ or $\textit{Round}_+(r_e,e)$. For instance, a expression $e = a \oplus b$ is replaced by a variable $r_e$, and the constraint $\textit{Round}_+(r_e, a+b)$ is added to the semantics. In the case of a compound expression $e = ab+c$, we introduce $e_1 = ab$, and we obtain $\textit{Round}_+(r_e, r_{e_1}+c) \land \textit{Round}(r_{e_1}, ab)$. If we know that the compiler uses a fused multiply-add operator, we can use $\textit{Round}(r_e,ab+c)$ instead.

\section{Complex control flow}
\label{part:control-flow}
We have so far assumed no procedure call, and at most one single loop. We shall see here how to deal with arbitrary control flow graphs and call graph structures.

\subsection{Loop Nests}
\label{part:noop_nests}
In Sec.~\ref{part:template_linear_constraints_invariants}, we have explained how to abstract a single fixed point. The method can be applied to multiple nested fixed points by replacing the inner fixed point by its abstraction. For instance, assume the rate limiter of Sec.~\ref{part:rate_limiter} is placed inside a larger loop. One may replace it by its abstraction:

\begin{verbatim}
if (e1max > e3max) {
  s1max = e1max;
} else {
  s1max = e3max;
} 
assume(s1 <= s1max);
/* and similar for s1min */
\end{verbatim}

Alternatively, we can extend our framework to an arbitrary control flow graph with nested loops, the semantics of which is expressed as a single fixed point. We may use the same method as proposed by \citet[\S2]{Gulwani_PLDI08} and other authors. First, a \emph{cut set} of program locations is identified; any cycle in the control flow graph must go through at least one program point in the cut set. In widening-based fixed point approximations, one classically applies widening at each point in the cut set. A simple method for choosing a cut set is to include all targets of back edges in a depth-first traversal of the control-flow graph, starting from the start node; in the case of structured program, this amounts to choosing the head node of each loop. This is not necessarily the best choice with respect to precision, though~\cite[\S2.3]{Gulwani_PLDI08}; \citet[Sec.~3.6]{Bourdoncle_PhD} discusses methods for choosing such as cut-set.

To each point in the cut set we associate an element in the abstract domain, parameterized by a number of variables. The values of these variables for all points in the cut-set defines an invariant candidate. Since paths between elements of the cut sets cannot contain a cycle, their denotational semantics can be expressed simply by an existentially quantified formula. Possible paths between each source and destination elements in the cut-set defined a stability condition (Formula~\ref{eqn:stability}). The conjunction of all these stability conditions defines acceptable inductive invariants. As above, the least inductive invariant is obtained by writing a minimization formula (Sec.~\ref{part:template_linear_constraints_invariants}).

Let us take a simple example:
\begin{verbatim}
i=0;
while(true) { /* A */
  if (choice()) {
    j=0;
    while(j < i) { /* B */
      /* something */
      j=j+1;
    }
    i=i+1;
    if (i==20) {
      i=0;
    }
  } else {
    /* something */
  }
}  
\end{verbatim}

We choose program points $A$ and $B$ as cut-set. At program point A, we look for an invariant of the form $I_A(i,j) \definedAs i_{\min,A} \leq i \leq i_{\max,A}$, and at program point B, for an invariant of the form $I_B(i,j) \definedAs i_{\min,B} \leq i \leq i_{\max,B} \land j_{\min} \leq j \leq j_{\max} \land \delta_{\min} \leq i-j \leq \delta_{\max}$ (a \emph{difference-bound} invariant). The (somewhat edited for brevity) stability formula is written:

\begin{multline}
\forall j~ I_A(0,j)\land \forall i \forall j~
((I_B(i,j)\land j\geq i\land (i+1\leq 19\lor\\
i+1=20\lor i+1\geq 21))\Rightarrow
\text{If}[i+1=20,I_A(0,j),I_A(i+1,j)])\land\\
\forall i \forall j~ (I_A(i,j)\Rightarrow
I_B(i,0))\land \forall i \forall j~((I_B(i,j)\land
j<i)\\
\Rightarrow I_B(i,j+1))
\end{multline}

Replacing $I_A$ and $I_B$ into this formula, then applying quantifier elimination, we obtain a formula defining all acceptable tuples $(i_{\min,A}, i_{\max,A}, i_{\min,B}, i_{\max,B}, j_{\min}, j_{\max}, \delta_{\min}, \delta_{\max})$. Optimal values are then obtained by further quantifier elimination: $i_{\min,A}=i_{\min,B}=j_{\min}=0$, $i_{\max,A}=i_{\max,B}=19$, $j_{\max}=20$, $\delta_{\min}=1$, $\delta_{\max}=19$.

The same example can be solved by replacing $20$ by another variable \texttt{n} as in Sec.~\ref{part:loop_example}.

\subsection{Procedures and Recursive Procedures}
We have so far considered abstractions of program blocks with respect to sets of program states. A program block is considered as a transformer from a state of input program states to the corresponding set of output program states. The analysis outputs a sound and optimal (in a certain way) abstract transformer, mapping an abstract set of input states to an abstract set of output states.

Assuming there are no recursive procedures, procedure calls can be easily dealt with. We can simply inline the procedure at the point of call, as done in e.g. \textsc{Astr\'ee} \cite{BlanchetCousotEtAl02-NJ,ASTREE_PLDI03,ASTREE_ESOP05}. Because inlining the concrete procedure may lead to code blowup, we may also inline its abstraction, considered as a nondeterministic program. Consider a complex procedure \texttt{P} with input variable \texttt{x} and output variable \texttt{x}. We abstract the procedure automatically with respect to the interval domain for the postcondition ($m_z \leq z \leq M_z$); suppose we obtain $M_z:=1000; m_z:=x$ then we can replace the function call by \verb@z <= 1000 && z >= x@. This is a form of \emph{modular interprocedural analysis}: considering the call graph, we can abstract the leaf procedures, then those calling the leaf procedures and so on. This method is however insufficient for dealing with recursive procedures.

In order to analyze recursive procedures, we need to abstract not sets of states, but sets of pairs of states, expressing the input-output relationships of procedures. In the case of recursive procedures, these relationships are the least solution of a system of equations.

To take a concrete example, let us consider McCarthy's famous ``91 function''~\cite{Manna_McCarthy_69,Manna_Pnueli_JACM70}, which, non-obviously, returns 91 for all inputs less than 101:
\begin{verbatim}
int M(int n) {
  if (n > 100) {
    return n-10;
  } else {
    return M(M(n+11));
  }
}
\end{verbatim}

The concrete semantics of that function is a relationship $R$ between its input $n$ and its output $r$. It is the least solution of
\begin{multline}\label{eqn:McCarthy91_stability}
R \supseteq \{ (n, r) \in \bbZ^2 \mid (n > 100 \land r=n-10) \lor\\
   (n \leq 100 \land \exists n_2 \in \bbZ
   (n+11, n_2) \in R \land (n_2, r) \in R) \}
\end{multline}

We look for a inductive invariant of the form $I \definedAs ((n \geq A) \land (r-n \geq \delta) \land (r-n \leq \Delta)) \lor ((n \leq B) \land (r=C))$, a non-convex domain (Sec.~\ref{part:non-convex}). By replacing $R$ by $I$ into inclusion~\ref{eqn:McCarthy91_stability}, and by universal quantification over $n,r,n_2$, we obtain the set of admissible parameters for invariants of this shape. By quantifier elimination, we obtain $(C=91) \land (\delta=\Delta=-10) \land (A=101) \land (B=100)$ within a fraction of a second using \textsc{Mjollnir} (see Sec.~\ref{part:experiments}).

In this case, there is a single acceptable inductive invariant of the suggested shape. In general, there may be parameters to optimize, as explained in Sec.~\ref{part:template_linear_constraints_invariants}. The result of this analysis is therefore a map from parameters defining sets of states to parameters defining sets of pairs of states (the abstraction of a transition relation). This abstract transition relation (a subset of $X \times Y$ where $X$ and $Y$ are the input and output state sets) can be transformed into an abstract transformer in $X^\sharp \rightarrow Y^\sharp$ as explained in Sec.~\ref{part:template_linear_constraints_transformers}. Such an interprocedural analysis may also be used to enhance the analysis of loops~\cite{DBLP:conf/cc/MartinAWF98}.

\section{Implementations and Experiments}
\label{part:experiments}
We have implemented the techniques of Sec.~\ref{part:template_linear_constraints_abstraction} in quantifier elimination packages, including  \textsc{Mathematica}%
\footnote{\url{http://www.wolfram.com/}}
and \textsc{Reduce}~3.8\footnote{\url{http://www.uni-koeln.de/REDUCE/}}
+ \textsc{Redlog}\footnote{\url{http://www.algebra.fim.uni-passau.de/~redlog/}}
in addition to our own package, \textsc{Mjollnir}~\cite{Monniaux_LPAR08}.%
\footnote{Source code and GNU/Linux/IA32 binaries of this implementation are available from \url{http://www-verimag.imag.fr/~monniaux/download/automatic_abstraction.zip}.}

As test cases, we took a library of operators for synchronous programming, having streams of floating-point values as input and outputs. These operators are written in a restricted subset of~C and take as much as 20 lines. A front-end based on \textsc{CIL}~\cite{Cil} converts them into formulas, then these formulas are processed and the corresponding abstract transfer functions are pretty-printed. Since for our application, it is important to bound numerical quantities, we chose the interval domain.

For instance, the rate limiter presented in Sec.~\ref{part:rate_limiter} was extracted from that library. Since this operator includes a memory (a variable whose value is retained from a call to the operator to the next one), its data-flow semantics is expressed using a fixed-point. When considered with real variables, the resulting expanded formula was approximately 1000 characters long, and with floating point variables approximately 8000 characters long. Despite the length of these formulas, they can be processed by \textsc{Mjollnir} in a matter of seconds. The result can then be saved once and for all.

Analyzers such as \textsc{Astr\'ee}~\cite{BlanchetCousotEtAl02-NJ,ASTREE_PLDI03,ASTREE_ESOP05} must have special knowledge about such operators, otherwise the analysis results are too coarse (for instance, the intervals do not get stabilized at all). The \textsc{Astr\'ee} development team therefore had to provide specialized, hand-written analyzes. In contrast, all linear floating-point operators in the library were analyzed within a fraction of a second using the method in the present article, assuming that floating-point values in the source code were real numbers. If one considered instead the abstraction of floating-point computations using real numbers from Sec.~\ref{part:float}, computation times did not exceed 17~seconds per operator; the formulas produced are considerably more complex than in the real case. Note that this computation is done once and for all for each operator; a static analyzer can therefore cache this information for further use and need not recompute abstractions for library functions or operators unless these functions are updated.

Our analyzer front-end currently cannot deal with non-numerical operations and data structures (pointers, records, and arrays). It is therefore not yet capable of directly dealing with the real control-command programs that e.g. \textsc{Astr\'ee} accepts, which do not consist purely of numerical operators. We plan to integrate our analysis method into a more generic analyzer. Alternatively, we plan to adapt a front-end for synchronous programming languages such as~\textsc{Simulink}, a tool widely used by control/command engineers.

The correctness of the methods described in this article does not rely on any particularity of the quantifier elimination procedure used, provided one also has symbolic computation procedures for e.g. putting formulas in disjunctive normal form and simplifying them. The difference between the various quantifier elimination and simplification procedures is efficiency; experiments showed that ours was vastly more efficient than the others tested for this kind of application. For instance, our implementation was able to complete the analysis of the rate limiter of Sec.~\ref{part:rate_limiter}, implemented over the reals, in 1.4~s, and in 17~s with the same example over floating-point numbers, while \textsc{Redlog} took 182~s for the former and could not finish the latter, and \textsc{Mathematica} could analyze neither (out-of-memory). On other examples, our quantifier elimination procedure is faster than the other ones, or can complete eliminations that the others cannot~\cite{Monniaux_LPAR08}.

\section{Related Works}
\label{part:related}
There is a sizeable amount of literature concerning relational numerical abstract domains; that is, domains that express constraints between numerical variables. Convex polyhedra were proposed in the 1970s~\cite{Halbwachs_PhD,CousotHalbwachs78}, and there have been since then many improvements to the technique; a bibliography was gathered by~\citet{PPL}. Algorithms on polyhedra are costly and thus a variety of domains intermediate between simple interval analysis and convex polyhedra were proposed~\cite{Mine_AST_WCRE01,Clariso_Cortadella_SAS2004,Sankaranarayana+others/05/Scalable}. All these domains compute invariants using a \emph{widening} operator~\cite{CousotCousot76-1,CousotHalbwachs78,CC92}. There is, however, no guarantee that the resulting invariant is the best representable in the abstract domain, even with the use of \emph{narrowing} iterations; this is one difference with our proposal, which computes the best representable inductive invariant.

Another difference is that these domains are designed to work with numerical values for the input constraints, thus the computation must be done for every value of the input constraints parameters. Using simple program transformations, they may also apply to symbolic input constraints (constraint parameters being taken as extra variables), but in general this will lead to bad results; for instance, the input-output relationship for the rate limiter of Sec.~\ref{part:rate_limiter} is not convex, while numerical abstract domains in the literature are convex. In comparison the algorithm in this article can be run once to obtain a \emph{formula} that gives the best invariant depending on the input constraints, allowing \emph{modular} analysis.

Several methods have been proposed to synthesize invariants without using widening operators~\cite{Colon_CAV03,Cousot05-VMCAI,Sankaranarayanan_SAS04}. In common with us, they express as constraints the conditions under which some parametric invariant shape truly is an invariant, then they use some resolution or simplification technique over those constraints. Again, these methods are designed for solving the problem for one given set of constraints on the inputs, as opposed to finding a relation between the output or fixed-point constraints and the input constraints. In some cases, the invariant may also not be minimal.

\citet{DBLP:conf/sas/BagnaraHMZ05,DBLP:journals/scp/BagnaraHRZ05} proposed improvements over the ``classical'' widenings on linear constraint domains~\cite{Halbwachs_PhD}. \citet{DBLP:conf/cav/GopanR06} introduced ``lookahead widenings'': standard widening-based analysis is applied to a sequence of syntactic restrictions of the original program, which ultimately converges to the whole programs; the idea is to distinguish phases or modes of operation in order to make the widening more precise.
\citet{DBLP:conf/sas/GonnordH06} have proposed \emph{acceleration} techniques for linear constraints. These do not replace widenings altogether, but they alleviate the need for some of the costly workarounds to the imprecision introduced by widenings, such as delayed widening~\cite[Sec.~7.1.3]{ASTREE_PLDI03}. These address a different problem from ours. On the one hand, neither improved widenings nor acceleration guarantee that the inductive invariant obtained at the end is the least one (indeed, they can yield the top element~$\top$).
\footnote{There exist \emph{exact acceleration} techniques but these rather apply to discrete automata.}
Furthermore, the invariant that these methods obtain is not parametric in the precondition, contrary to the one that our method obtains. On the other hand, improved widenings work regardless of the form of the transition relation, which our method constrains to be piecewise linear. Some of the cited methods operate on general polyhedra, while our method constrains the shape of the polyhedra that are found to a certain template.

\citet{GGTZ:07,Gawlitza_Seidl_ESOP07} proposed replacing the usual widening/narrowing iteration techniques by a \emph{policy iteration} (or \emph{strategy iteration}) approach. Their approach converges on a fixed point, but not necessarily the least one. Their idea is to replace computing the least fixed point of a complex abstract operator (the point-wise minimum of a family of simpler operators) by a sequence of least fixed point computations for these simple operators. Their technique anyway needs to compute these latter least fixed points, and it is possible that our method can help in that respect.

Techniques using quantifier elimination for generating nonlinear invariants for programs using nonlinear arithmetic have also been proposed \cite{Kapur_ACA04} and shown capable of producing optimal invariants parameterized by input constraints \cite{Monniaux_SAS07}. Quantifier elimination in the theory of real closed fields is, however, a very costly technique. Experimentally, the formulas generated by common implementations tend to grow huge (due to difficult simplifications) and both time and space requirements grow very fast with the number of variables. This is why we considered the linear case in the present article.

\citet{Gulwani_PLDI08} have also proposed a method for generating linear invariants over \emph{integer} variables, using a class of templates. The methods described in the present article can be applied to linear invariants over integer variables in two ways: either by abstracting them using rationals (as in examples in Sec.~\ref{part:loop_example}, \ref{part:noop_nests}), either by replacing quantifier elimination over rational linear arithmetic by quantifier elimination over linear integer arithmetic, also known as Presburger arithmetic. Quantifier elimination over Presburger arithmetic is however very expensive \cite{Fischer_Rabin_exponential_74}. \citeauthor{Gulwani_PLDI08} instead chose to first consider integer variables as rationals, so as to be able to compute over rational convex polyhedra, then bound variables and constraint parameters so as to model them as finite bit vectors, finally obtaining a problem amenable to SAT~solving. Program variables \emph{are} finite bit vectors in most industrial programming languages, and parameters to useful invariants over integer variables are often small, thus their approach seems justified. We do not see, however, how their method could be applied to programs operating over real or floating-point variables, which are the main motivation for the present article.

The idea of producing procedure summaries \cite{Sharir_Pnueli_81} as formulas mapping input bounds to output bounds is not new. \citet{Rugina_Rinard_05}, in the context of pointer analysis (with pointers considered as a base plus an integer offset), proposed a reduction to linear programming. This reduction step, while sound, introduces an imprecision that is difficult to measure in advance; our method, in contrast, is guaranteed to be ``optimal'' in a certain sense. \citeauthor{Rugina_Rinard_05}'s method, however, allows some nonlinear constructs in the program to be analyzed. \citet{DBLP:conf/cc/MartinAWF98} proposed applying interprocedural analysis to loops.

\citet{Seidl_ESOP07} also produce procedure summaries as numerical constraints. Our procedure summaries are implementations of the corresponding abstract transformer over some abstract domain, while theirs outputs a relationship between input and output concrete values. Their analysis considers a \emph{convex} set of concrete input-output relationships, expressed as a \emph{simplices}, a restricted class of convex polyhedra. This restriction trades precision for speed: the generator and constraint representations of simplices have approximately the same size, while in general polyhedra exponential blowup can occur. Tests by arbitrary linear constraints cannot be adequately represented within this framework. \citet[Sec.~4]{Seidl_ESOP07} propose deferring those constraints using auxiliary variables; this, however, loses some precision. Their analysis and ours are therefore incomparable, since they make different choices between precision and efficiency.

\citet{Reps_CAV05} proposed an interprocedural analysis of numerical properties of functions using weighted pushdown automata. The ``weights'' are taken in a finite height abstract domain, while the domains we consider have infinite height.

In earlier works we have proposed a method for obtaining input-output relationships of digital linear filters with memories, taking into account the effects of floating-point computations~\cite{Monniaux_CAV05}. This method computes an exact relationship between bounds on the input and bounds on the output, without the need for an abstract domain for expressing the local invariant; as such, for this class of problems, it is more precise than the method from this article. This technique, however, cannot be easily generalized to cases where the operator block contains tests.

\section{Conclusion and Future Work}
Writing static analyzers by hand has long been found tedious and error-prone. One may of course prove an existing analyzer correct through assisted proof techniques, which removes the possibility of soundness mistakes, at the expense of much increased tediousness. In this article, we proposed instead effective methods to synthesize abstract domains by automatic techniques. The advantages are twofold: new domains can be created much more easily, since no programming is involved; a single procedure, testable on independent examples, needs be written and possibly formally proved correct. To our knowledge, this is the first effective proposal for generating numerical abstract domains automatically, and one of the few methods for generating numerical summaries. Also, it is also the only method so far for computing summaries of \emph{floating-point} functions.

We have shown that floating-point computations could be safely abstracted using our method. The formulas produced are however fairly complex in this case, and we suspect that further over-approximation could dramatically reduce their size. There is also nowadays significant interest in automatizing, at least partially, the tedious proofs that computer arithmetic experts do and we think that the kind of methods described in this article could help in that respect.

We have so far experimented with small examples, because the original goal of this work was the automatic, on-the-fly, synthesis of abstract transfer functions for small sequences of code that could be more precise than the usual composition of abstract of individual instructions, and less tedious for the analysis designer than the method of pattern-matching the code for ``known'' operators with known mathematical properties. A further goal is the precise analysis of longer sequences, including integer and Boolean computations. We have shown in Sec.~\ref{part:partitioning} how it was possible to partition the state space and abstract each region of the state-space separately; but naive partitioning according to $n$ Booleans leads to $2^n$ regions, which can be unbearably costly and is unneeded in most cases. We think that automatic refinement and partitioning techniques \cite{jeannet03a} could be developed in that respect.

\phantomsection\addcontentsline{toc}{section}{References}

\bibliographystyle{plainnat}

\begin{thebibliography}{49}
\providecommand{\natexlab}[1]{#1}
\providecommand{\url}[1]{\texttt{#1}}
\expandafter\ifx\csname urlstyle\endcsname\relax
  \providecommand{\doi}[1]{doi: #1}\else
  \providecommand{\doi}{doi: \begingroup \urlstyle{rm}\Url}\fi

\bibitem[Bagnara et~al.(2005{\natexlab{a}})Bagnara, Hill, Mazzi, and
  Zaffanella]{DBLP:conf/sas/BagnaraHMZ05}
Roberto Bagnara, Patricia~M. Hill, Elena Mazzi, and Enea Zaffanella.
\newblock \href{http://arxiv.org/abs/cs.PL/0412043}{Widening operators for weakly-relational numeric abstractions}.
\newblock In \emph{Static Analysis (SAS)}, volume 3672 of \emph{LNCS}, pages
  3--18. Springer, 2005{\natexlab{a}}.
\newblock \doi{10.1007/11547662\_3}.

\bibitem[Bagnara et~al.(2005{\natexlab{b}})Bagnara, Hill, Ricci, and
  Zaffanella]{DBLP:journals/scp/BagnaraHRZ05}
Roberto Bagnara, Patricia~M. Hill, Elisa Ricci, and Enea Zaffanella.
\newblock \href{http://www.cs.unipr.it/Publications/Abstracts/Q399}{Precise widening operators for convex polyhedra}.
\newblock \emph{Sci. Comput. Program.}, 58\penalty0 (1-2):\penalty0 28--56,
  2005{\natexlab{b}}.
\newblock \doi{10.1016/j.scico.2005.02.003}.

\bibitem[Bagnara et~al.(2006)Bagnara, Hill, and Zaffanella]{PPL}
Roberto Bagnara, Patricia~M. Hill, and Enea Zaffanella.
\newblock \emph{The Parma Polyhedra Library, version 0.9}, 2006.
\newblock URL \url{http://www.cs.unipr.it/ppl}.

\bibitem[Balakrishnan and Reps(2004)]{DBLP:conf/cc/BalakrishnanR04}
Gogul Balakrishnan and Thomas Reps.
\newblock \href{http://www.cs.wisc.edu/wpis/papers/cc04.pdf}{Analyzing memory accesses in x86 executables}.
\newblock In \emph{Compiler Construction (CC)}, volume 2985 of \emph{LNCS},
  pages 5--23. Springer, 2004.
\newblock \doi{10.1007/b95956}.

\bibitem[Blanchet et~al.(2002)Blanchet, Cousot, Cousot, Feret, Mauborgne,
  Min\'e, Monniaux, and Rival]{BlanchetCousotEtAl02-NJ}
Bruno Blanchet, Patrick Cousot, Radhia Cousot, J\'er\^ome Feret, Laurent
  Mauborgne, Antoine Min\'e, David Monniaux, and Xavier Rival.
\newblock \href{http://www.di.ens.fr/~rival/njones.pdf}{Design and implementation of a special-purpose static program
  analyzer for safety-critical real-time embedded software}.
\newblock In \emph{The Essence of Computation: Complexity, Analysis,
  Transformation}, number 2566 in LNCS, pages 85--108. Springer, 2002.
\newblock \doi{10.1007/3-540-36377-7\_5}.

\bibitem[Blanchet et~al.(2003)Blanchet, Cousot, Cousot, Feret, Mauborgne,
  Min\'e, Monniaux, and Rival]{ASTREE_PLDI03}
Bruno Blanchet, Patrick Cousot, Radhia Cousot, J\'er\^ome Feret, Laurent
  Mauborgne, Antoine Min\'e, David Monniaux, and Xavier Rival.
\newblock \href{http://arxiv.org/abs/cs.PL/0701193}{A static analyzer for large safety-critical software}.
\newblock In \emph{Programming Language Design and Implementation (PLDI)},
  pages 196--207. ACM, 2003.
\newblock \doi{10.1145/781131.781153}.

\bibitem[Bourdoncle(1992)]{Bourdoncle_PhD}
Fran\c{c}ois Bourdoncle.
\newblock \href{http://www.exalead.com/Francois.Bourdoncle/these.html}{\emph{S\'emantique des langages imp\'eratifs d'ordre sup\'erieur et
  interpr\'etation abstraite}}.
\newblock PhD thesis, \'Ecole polytechnique, Palaiseau, 1992.

\bibitem[Bradley and Manna(2007)]{BradleyManna07}
Aaron~R. Bradley and Zohar Manna.
\newblock \emph{The Calculus of Computation: Decision Procedures with
  Applications to Verification}.
\newblock Springer, October 2007.

\bibitem[Caspi et~al.(1987)Caspi, Pilaud, Halbwachs, and Plaice]{LUSTRE}
Paul Caspi, Daniel Pilaud, Nicolas Halbwachs, and John~A. Plaice.
\newblock {LUSTRE}: a declarative language for real-time programming.
\newblock In \emph{Principles of Programming Languages (POPL)}, pages 178--188.
  ACM, 1987.
\newblock \doi{10.1145/41625.41641}.

\bibitem[Caspi et~al.(2003)Caspi, Curic, Maignan, Sofronis, Tripakis, and
  Niebert]{Caspi_et_al_2003}
Paul Caspi, Adrian Curic, Aude Maignan, Christos Sofronis, Stavros Tripakis,
  and Peter Niebert.
\newblock From {S}imulink to {S}cade/{L}ustre to {TTA}: a layered approach for
  distributed embedded applications.
\newblock \emph{SIGPLAN notices}, 38\penalty0 (7):\penalty0 153--162, 2003.
\newblock \doi{10.1145/780731.780754}.

\bibitem[CAV05()]{CAV_2005}
CAV05.
\newblock \emph{Computer Aided Verification (CAV)}, number 4590 in LNCS, 2005.
  Springer.
\newblock \doi{10.1007/b138445}.

\bibitem[Claris\'o and Cortadella(2004)]{Clariso_Cortadella_SAS2004}
Robert Claris\'o and Jordi Cortadella.
\newblock The octahedron abstract domain.
\newblock In \emph{Static Analysis (SAS)}, number 3148 in LNCS, pages 312--327.
  Springer, 2004.

\bibitem[Colon et~al.(2003)Colon, Sankaranarayanan, and Sipma]{Colon_CAV03}
Michael Colon, Sriram Sankaranarayanan, and Henny Sipma.
\newblock Linear invariant generation using non-linear constraint solving.
\newblock In \emph{Computer Aided Verification (CAV)}, number 2725 in LNCS,
  pages 420--433. Springer, 2003.

\bibitem[Cousot(2005)]{Cousot05-VMCAI}
Patrick Cousot.
\newblock \href{http://www.di.ens.fr/~cousot/COUSOTpapers/VMCAI05.shtml}{%
  Proving program invariance and termination by parametric abstraction,
  lagrangian relaxation and semidefinite programming}.
\newblock In \citet{VMCAI_2005}, pages 1--24.
\newblock \doi{10.1007/b105073}.

\bibitem[Cousot and Cousot(1992)]{CC92}
Patrick Cousot and Radhia Cousot.
\newblock \href{http://www.di.ens.fr/~cousot/COUSOTpapers/JLP92.shtml}{%
  Abstract interpretation and application to logic programs}.
\newblock \emph{J. of Logic Programming}, 13\penalty0 (2--3):\penalty0
  103--179, 1992.

\bibitem[Cousot and Cousot(1976)]{CousotCousot76-1}
Patrick Cousot and Radhia Cousot.
\newblock \href{http://www.di.ens.fr/~cousot/COUSOTpapers/ISOP76.shtml}{%
  Static determination of dynamic properties of programs}.
\newblock In \emph{Proceedings of the Second International Symposium on
  Programming}, pages 106--130. Dunod, Paris, France, 1976.

\bibitem[Cousot and Halbwachs(1978)]{CousotHalbwachs78}
Patrick Cousot and Nicolas Halbwachs.
\newblock \href{http://citeseer.ist.psu.edu/cousot78automatic.html}{%
  Automatic discovery of linear restraints among variables of a program}.
\newblock In \emph{Principles of Programming Languages (POPL)}, pages 84--96.
  ACM, 1978.
\newblock \doi{10.1145/512760.512770}.

\bibitem[Cousot et~al.(2005)Cousot, Cousot, Feret, Mauborgne, Min\'e, Monniaux,
  and Rival]{ASTREE_ESOP05}
Patrick Cousot, Radhia Cousot, J\'er\^ome Feret, Laurent Mauborgne, Antoine
  Min\'e, David Monniaux, and Xavier Rival.
\newblock \href{http://hal.archives-ouvertes.fr/hal-00084293/en/}{The {ASTR\'EE} analyzer}.
\newblock In \emph{Programming Languages and Systems (ESOP)}, number 3444 in
  LNCS, pages 21--30, 2005.

\bibitem[ESOP07()]{ESOP_2007}
ESOP07.
\newblock \emph{Programming Languages and Systems (ESOP)}, volume 4421 of
  \emph{LNCS}, 2007. Springer.
\newblock \doi{10.1007/978-3-540-71316-6}.

\bibitem[Ferrante and Rackoff(1975)]{FerranteRackoff75}
Jeanne Ferrante and Charles Rackoff.
\newblock A decision procedure for the first order theory of real addition with
  order.
\newblock \emph{{SIAM} Journal of Computation}, 4\penalty0 (1):\penalty0
  69--76, March 1975.

\bibitem[Fischer and Rabin(1974)]{Fischer_Rabin_exponential_74}
Michael~J. Fischer and Michael~O. Rabin.
\newblock \href{http://www.lcs.mit.edu/publications/pubs/ps/MIT-LCS-TM-043.ps}{Super-exponential complexity of Presburger arithmetic}.
\newblock In \emph{Complexity of Computation}, number~7 in SIAM--AMS
  proceedings, pages 27--42. American Mathematical Society, 1974.

\bibitem[Gaubert et~al.(2007)Gaubert, Goubault, Taly, and Zennou]{GGTZ:07}
St\'ephane Gaubert, \'Eric Goubault, Ankur Taly, and Sarah Zennou.
\newblock Static analysis by policy iteration on relational domains.
\newblock In \citet{ESOP_2007}, pages 237--252.
\newblock \doi{10.1007/978-3-540-71316-6}.

\bibitem[Gawlitza and Seidl(2007)]{Gawlitza_Seidl_ESOP07}
Thomas Gawlitza and Helmut Seidl.
\newblock Precise fixpoint computation through strategy iteration.
\newblock In \citet{ESOP_2007}, pages 300--315.
\newblock \doi{10.1007/978-3-540-71316-6\_21}.

\bibitem[Gonnord and Halbwachs(2006)]{DBLP:conf/sas/GonnordH06}
Laure Gonnord and Nicolas Halbwachs.
\newblock \href{http://hal.archives-ouvertes.fr/hal-00189614/en/}{Combining widening and acceleration in linear relation analysis}.
\newblock In \emph{Static Analysis (SAS)}, volume 4134 of \emph{LNCS}, pages
  144--160. Springer, 2006.
\newblock \doi{10.1007/11823230\_10}.

\bibitem[Gopan and Reps(2006)]{DBLP:conf/cav/GopanR06}
Denis Gopan and Thomas~W. Reps.
\newblock \href{http://www.cs.wisc.edu/wpis/papers/cav06-widening.pdf}{Lookahead widening}.
\newblock In \emph{Computer Aided Verification (CAV)}, volume 4144 of
  \emph{LNCS}, pages 452--466. Springer, 2006.
\newblock \doi{10.1007/11817963\_41}.

\bibitem[Gopan and Reps(2007)]{DBLP:conf/cav/GopanR07}
Denis Gopan and Thomas~W. Reps.
\newblock \href{http://www.cs.wisc.edu/wpis/papers/cav07.summarization.pdf}{Low-level library analysis and summarization}.
\newblock In \emph{Computer Aided Verification (CAV)}, volume 4590 of
  \emph{LNCS}, pages 68--81. Springer, 2007.
\newblock \doi{10.1007/978-3-540-73368-3\_10}.

\bibitem[Gulwani et~al.(2008)Gulwani, Srivastava, and
  Venkatesan]{Gulwani_PLDI08}
Sumit Gulwani, Saurabh Srivastava, and Ramarathnam Venkatesan.
\newblock \href{http://research.microsoft.com/users/sumitg/pubs/pldi08_cs.ps}{Program analysis as constraint solving}.
\newblock In \emph{Programming Language Design and Implementation (PLDI)}. ACM,
  2008.
\newblock \doi{10.1145/1375581.1375616}.

\bibitem[Halbwachs(1979)]{Halbwachs_PhD}
Nicolas Halbwachs.
\newblock \emph{D\'etermination automatique de relations lin\'eaires
  v\'erifi\'ees par les variables d'un programme}.
\newblock PhD thesis, Universit\'e scientifique et m\'edicale de Grenoble,
  1979.

\bibitem[{IEE}(1985)]{IEEE-754}
\emph{IEEE standard for Binary floating-point arithmetic for microprocessor
  systems}.
\newblock {IEEE}, 1985.
\newblock ANSI/IEEE Std 754-1985.

\bibitem[Jeannet(2003)]{jeannet03a}
Bertrand Jeannet.
\newblock Dynamic partitioning in linear relation analysis. application to the
  verification of reactive systems.
\newblock \emph{Formal Methods in System Design}, 23\penalty0 (1):\penalty0
  5--37, July 2003.

\bibitem[Kapur(2004)]{Kapur_ACA04}
Deepak Kapur.
\newblock \href{http://www.cs.unm.edu/~kapur/mypapers/aca2004.pdf}{%
  Automatically generating loop invariants using quantifier elimination}.
\newblock In \emph{ACA (Applications of Computer Algebra)}, 2004.

\bibitem[Lal et~al.(2005)Lal, Balakrishnan, and Reps]{Reps_CAV05}
Akash Lal, Gogul Balakrishnan, and Thomas Reps.
\newblock \href{http://www.cs.wisc.edu/wpis/papers/cav05-ewpds.pdf}{%
  Extended weighted pushdown systems}.
\newblock In \citet{CAV_2005}, pages 343--357.
\newblock \doi{10.1007/11817963\_32}.

\bibitem[Logozzo and F{\"a}hndrich(2008)]{DBLP:conf/cc/LogozzoF08}
Francesco Logozzo and Manuel F{\"a}hndrich.
\newblock \href{http://research.microsoft.com/~logozzo/publications/papers/cc08.pdf}{On the relative completeness of bytecode analysis versus source code
  analysis}.
\newblock In \emph{Compiler Construction (CC)}, volume 4959 of \emph{LNCS},
  pages 197--212. Springer, 2008.
\newblock \doi{10.1007/978-3-540-78791-4\_14}.

\bibitem[Manna and McCarthy(1969)]{Manna_McCarthy_69}
Zohar Manna and John McCarthy.
\newblock Properties of programs and partial function logic.
\newblock In \emph{Machine Intelligence}, 5, pages 27--38. Edinburgh University
  Press, 1969.

\bibitem[Manna and Pnueli(1970)]{Manna_Pnueli_JACM70}
Zohar Manna and Amir Pnueli.
\newblock Formalization of properties of functional programs.
\newblock \emph{J. ACM}, 17\penalty0 (3):\penalty0 555--569, 1970.
\newblock \doi{10.1145/321592.321606}.

\bibitem[Martin et~al.(1998)Martin, Alt, Wilhelm, and
  Ferdinand]{DBLP:conf/cc/MartinAWF98}
Florian Martin, Martin Alt, Reinhard Wilhelm, and Christian Ferdinand.
\newblock Analysis of loops.
\newblock In \emph{Compiler Construction (CC)}, volume 1383 of \emph{LNCS},
  pages 80--94. Springer, 1998.

\bibitem[Min\'e(2001)]{Mine_AST_WCRE01}
Antoine Min\'e.
\newblock \href{http://www.di.ens.fr/~mine/publi/article-mine-ast01.pdf}{%
  The octagon abstract domain}.
\newblock In \emph{Reverse Engineering (WCRE)}, pages 310--319. IEEE, 2001.
\newblock \doi{10.1109/WCRE.2001.957836}.

\bibitem[Min\'e(2004)]{Mine_ESOP04}
Antoine Min\'e.
\newblock \href{http://arxiv.org/abs/cs.PF/0703077}{%
  Relational abstract domains for the detection of floating-point
  run-time errors}.
\newblock In \emph{Programming Languages and Systems (ESOP)}, volume 2986 of
  \emph{LNCS}, pages 3--17. Springer, 2004.

\bibitem[Monniaux(2005)]{Monniaux_CAV05}
David Monniaux.
\newblock \href{http://hal.archives-ouvertes.fr/hal-00084291/en/}{Compositional analysis of floating-point linear numerical filters}.
\newblock In \citet{CAV_2005}, pages 199--212.
\newblock \doi{10.1007/b138445}.

\bibitem[Monniaux(2008{\natexlab{a}})]{Monniaux_LPAR08}
David Monniaux.
\newblock \href{http://hal.archives-ouvertes.fr/hal-00262312/en/}{%
  A quantifier elimination algorithm for linear real arithmetic}.
\newblock In \emph{LPAR (Logic for Programming, Artificial Intelligence, and
  Reasoning)}, LNCS. Springer, 2008{\natexlab{a}}.

\bibitem[Monniaux(2007)]{Monniaux_SAS07}
David Monniaux.
\newblock \href{http://hal.archives-ouvertes.fr/hal-00148608/en/}{%
  Optimal abstraction on real-valued programs}.
\newblock In \emph{Static analysis (SAS)}, number 4634 in LNCS, pages 104--120.
  Springer, 2007.
\newblock \doi{10.1007/978-3-540-74061-2\_7}.

\bibitem[Monniaux(2008{\natexlab{b}})]{Monniaux_TOPLAS08}
David Monniaux.
\newblock \href{http://hal.archives-ouvertes.fr/hal-00128124/en/}{%
  The pitfalls of verifying floating-point computations}.
\newblock \emph{ACM Transactions on programming languages and systems},
  30\penalty0 (3):\penalty0 12, 2008{\natexlab{b}}.
\newblock \doi{10.1145/1353445.1353446}.

\bibitem[Necula et~al.(2002)Necula, McPeak, Rahul, and Weimer]{Cil}
George~C. Necula, Scott McPeak, Shree~P. Rahul, and Westley Weimer.
\newblock \href{http://www.cs.berkeley.edu/~necula/Papers/cil_cc02.pdf}{%
  {CIL}: Intermediate language and tools for analysis and
  transformation of {C} programs}.
\newblock In \emph{Compiler Construction (CC)}, volume 2304 of \emph{LNCS},
  pages 209--265. Springer, 2002.
\newblock \doi{10.1007/3-540-45937-5\_16}.

\bibitem[Rugina and Rinard(2005)]{Rugina_Rinard_05}
Radu Rugina and Martin Rinard.
\newblock \href{http://www.cag.lcs.mit.edu/~rinard/paper/toplas05SymbolicBoundsAnalysis.pdf}{Symbolic bounds analysis for pointers, array indices, and accessed
  memory regions}.
\newblock \emph{ACM Trans. on Programming Languages and Systems (TOPLAS)},
  27\penalty0 (2):\penalty0 185--235, 2005.
\newblock \doi{10.1145/349299.349325}.

\bibitem[Sankaranarayanan et~al.(2004)Sankaranarayanan, Sipma, and
  Manna]{Sankaranarayanan_SAS04}
Sriram Sankaranarayanan, Henny Sipma, and Zohar Manna.
\newblock \href{http://www-step.stanford.edu/papers/sas04.html}{Constraint-based linear-relations analysis}.
\newblock In \emph{SAS}, number 3148 in LNCS, pages 53--68. Springer, 2004.

\bibitem[Sankaranarayanan et~al.(2005)Sankaranarayanan, Sipma, and
  Manna]{Sankaranarayana+others/05/Scalable}
Sriram Sankaranarayanan, Henny Sipma, and Zohar Manna.
\newblock \href{http://www-step.stanford.edu/papers/svmcai05.html}{Scalable analysis of linear systems using mathematical programming}.
\newblock In \citet{VMCAI_2005}, pages 21--47.
\newblock \doi{10.1007/b105073}.

\bibitem[Seidl et~al.(2007)Seidl, Flexeder, and Petter]{Seidl_ESOP07}
Helmut Seidl, Andrea Flexeder, and Michael Petter.
\newblock Interprocedurally analysing linear inequality relations.
\newblock In \citet{ESOP_2007}, pages 284--299.
\newblock \doi{10.1007/978-3-540-71316-6\_20}.

\bibitem[Sharir and Pnueli(1981)]{Sharir_Pnueli_81}
Micha Sharir and Amir Pnueli.
\newblock Two approaches to inter-procedural data-flow analysis.
\newblock In \emph{Program Flow Analysis: Theory and Application}.
  Prentice-Hall, 1981.

\bibitem[VMCAI05()]{VMCAI_2005}
VMCAI05.
\newblock \emph{Verification, Model Checking and Abstract Interpretation
  (VMCAI)}, number 3385 in LNCS, 2005. Springer.
\newblock \doi{10.1007/b105073}.

\end{thebibliography}

\end{document}